\documentclass[conference]{IEEEtran}
\usepackage[final]{graphicx}
\usepackage{subfigure}
\usepackage{float}
\usepackage{amsmath}
\usepackage{cases}
\usepackage{color}
\usepackage{colortbl}
\usepackage{authblk}
\usepackage{algorithm}
\usepackage{algpseudocode}
\usepackage{amsmath}
\usepackage{amssymb}
\usepackage{booktabs}
\usepackage{setspace}
\usepackage{array}
\makeatletter
\renewcommand{\maketag@@@}[1]{\hbox{\m@th\normalsize\normalfont#1}}%
\makeatother
\usepackage{stfloats}
\usepackage{cite}
\usepackage{makecell}
\usepackage{multirow}

\def\BibTeX{{\rm B\kern-.05em{\sc i\kern-.025em b}\kern-.08em
    T\kern-.1667em\lower.7ex\hbox{E}\kern-.125emX}}
    
\ifCLASSINFOpdf
\else
\fi

\usepackage{caption}
\usepackage{mathtools}

\begin{document}
\title{Beyond Single-Band: Analysis and Resource Allocation for Multi-band ISAC Systems}
\author[1]{Haotian Liu}
\author[1]{Zhiqing Wei}
\author[2]{Xingwang Li}
\author[1]{Yunxin Geng}
\author[1]{Qixun Zhang}
\author[1]{Zhiyong Feng}
\affil[1]{Beijing University of Posts and Telecommunications, Beijing 100876, China}
\affil[2]{Henan Polytechnic University, Jiaozuo 454000, China}
\affil[1]{Email:\{haotian\_liu, weizhiqing, gengyx, zhangqixun, fengzy\} @bupt.edu.cn}
\affil[2]{Email:lixingwangbupt@gmail.com}

\maketitle

\begin{abstract} 
Integrated sensing and communication (ISAC) has emerged as a pivotal technology for sixth-generation wireless networks to empower high-precision sensing. 
The demand for superior sensing resolution and the reality of spectrum fragmentation have driven the research of multi-band ISAC.
Multi-band ISAC provides frequency diversity through independent observations across disparate bands, mitigating sensing performance fluctuations caused by  frequency-selective radar cross-section compared to single-band counterparts.
In this paper, we propose a framework for analytical performance characterization and resource optimization in multi-band ISAC systems.
Specifically, analytical closed-form detection and false alarm probabilities for multi-band OFDM signals are derived, providing a theoretical foundation for subsequent resource allocation. 
Then, a joint power and time-frequency resource allocation scheme is developed and solved via a proposed ADMM-based algorithm to maximize detection performance.
Numerical results validate the accuracy of the closed-form derivations and demonstrate the superior robustness of multi-band signals. 
Notably, the proposed optimization scheme achieves an 18 dB detection gain over traditional single-band baselines at a 90\% detection probability.
\end{abstract}

\IEEEpeerreviewmaketitle

\section{Introduction}
%【第一段】6G网络旨在赋能无线网络感知能力，支持那些依赖高精度环境感知的新兴应用，如自动驾驶、低空经济、数字孪生。集成传感与通信作为支撑这一愿景的关键技术，吸引了学术界和工业界的广泛关注【】。
%不幸的是，传感性能受限于OFDM信号的带宽，而实际通信标准中可供感知使用的频谱始终是不连续的【OFDM-BASED MULTIBAND SENSING FOR ISAC】。
%受频段碎片化和高精度感知需求的影响，基于载波聚合技术的多频段ISAC研究，利用非连续、跨度大的异构频段（如毫米波与sub-6G）进行联合感知，已成为提升系统整体感知能力的关键路径。
Sixth-generation (6G) networks are expected to support high-precision environmental sensing for applications such as autonomous driving, the low-altitude economy, and digital twins~\cite{Fanliu_2022,wan2024}, motivating growing interest in integrated sensing and communication (ISAC)~\cite{Fanliu_2022,wan2024,wei2024deep,Li2025,Luo2024}.
However, sensing performance is limited by signal bandwidth, while the spectrum available in practical communication standards is fragmented~\cite{wan2024}.
Driven by the demand for high-precision sensing in weak-target scenarios, multi-band ISAC leveraging carrier aggregation (CA) has emerged as a pivotal approach~\cite{wei2024deep}. 
By facilitating joint sensing across non-contiguous, wide-span heterogeneous bands (e.g., sub-6 GHz and millimeter-wave), this paradigm significantly bolsters the overall sensing capability~\cite{wan2024,Li2025}.

%【第二段：说多频段相比于单频段的优势和动机】相比单频段OFDM感知信号，多频段OFDM感知信号的优势之一在于通过多频点观测获得频率分集增益【Multi-band composite detection and recognition of aerial infrared point targets】。这种特性能够补偿频段衰落与随机RCS波动，显著增强弱目标探测的鲁棒性。然而，由于多频段的参数异构性，以及目标散射特性的频段差异，解析表征多频段OFDM感知信号的探测性能并实现灵活的多维资源优化，是充分释放多频段感知潜力的关键【Multi-Node Multi-Band Cooperative Integrated Sensing and Communications: State-of-the-Art, Challenges and Opportunities】。
A key advantage of multi-band OFDM over single-band signals is the frequency diversity gain derived from multi-band independent observations~\cite{Tian2018}.
Such diversity effectively mitigates the impact of frequency-selective radar cross sections (RCS) and low echo energy of weak targets~\cite{Jung2011}.
However, the inherent parameter heterogeneity and diverse target RCS complicate the sensing process~\cite{liu2024carrier}, necessitating an analytical characterization of detection performance and the joint optimization of multi-dimensional resources to fully exploit the potential of multi-band sensing.

%【第三段】现有的多频段ISAC研究主要侧重于克拉美罗界（CRB）分析[我的两篇]、分辨率评估[VTM论文]以及感知信号处理算法设计[我的两篇]，而对于目标检测性能的解析表征及多维资源协同优化仍缺乏深入探讨。传统的多波段雷达探测已经成熟，但由于多波段OFDM信号的多种路径损耗、噪声和参数所导致的i.i.d.特性，阻碍了对探测概率推导出易于处理的封闭式表达式。其次，缺乏面向最大化检测性能的多频段ISAC系统的多维资源分配方案。
Existing multi-band ISAC research primarily emphasizes Cramer-Rao bound analysis~\cite{liu2024carrier,wei2023carrier}, resolution assessment~\cite{wan2024}, and signal processing algorithm design~\cite{liu2024carrier,wei2023carrier,Pegoraro2024}, while the analytical characterization of detection performance and multi-dimensional resource optimization remain under-explored. 
Although multi-band detection is well-established in traditional radar~\cite{Tian2018}, the independent non-identically distributed (i.n.i.d.) characteristics caused by various path losses, noise, and parameters of multi-band OFDM signals hinder the derivation of tractable closed-form expressions for detection probability.
Additionally, there is a lack of multi-dimensional resource allocation schemes for multi-band ISAC systems aimed at maximizing detection performance.

%【第四段-贡献】
% 综上所述，目前针对多频段 ISAC 系统的性能分析与资源优化研究仍显不足。本文针对具有频率选择性RCS的弱目标探测场景，建立了探测性能建模与资源协同优化框架。具体而言，我们基于广义似然比准则，推导出了多频段 OFDM 信号检测概率与虚警概率的解析闭式解。在此基础上，本文进一步提出了一种联合发射功率和时频资源的分配方案，并设计了基于 ADMM 的高效求解算法以最大化系统的检测性能。仿真结果验证了理论闭式解的准确性以及多频段信号相比于单频段信号在弱目标探测上的鲁棒性。结果表明，在 90% 的检测概率下，所提优化方案相比传统单频段信号可获得 18 dB 的探测增益。
Overall, existing research on the performance analysis and resource optimization of multi-band ISAC remains limited. 
In this paper, we develop a framework for analytical performance characterization and resource optimization in multi-band ISAC systems. 
Specifically, closed-form expressions for the detection and false alarm probabilities of multi-band OFDM signals are derived via the generalized likelihood ratio test (GLRT). 
Then, a joint power and time-frequency resource allocation scheme is proposed, along with an ADMM-based algorithm to maximize system detection performance. 
Simulations validate the accuracy of the analytical derivations and demonstrate the robust sensing performance of multi-band signals compared to single-band baselines. 
Notably, the proposed optimization scheme achieves an 18 dB detection gain over traditional single-band signals at a 90\% detection probability.

The remainder of this paper is structured as follows. Section \ref{se2} outlines the system model. Section \ref{se3} derives the detection performance. Section \ref{se4} presents the joint optimization scheme. Section \ref{se5} details the simulation results, and Section \ref{se6} summarizes the paper.

\textit{Notations:} $\{\cdot\}$ stands for a set of various index values.  
Black bold letters represent matrices or vectors.
$\mathbb{C}$ and $\mathbb{R}$ denote the sets of complex and real numbers, respectively. 
$\left[\cdot\right]^{*}$, $\approx$, $\mathbb{E}[\cdot]$, $\text{Exp}(\cdot)$, $\mathcal{F}^{-1}$, and $|\cdot|$ stand for the conjugate operator, approximate equal operator, expectation operator, exponential distribution, inverse Fourier transform operator, and absolute operator, respectively. 
% A complex Gaussian random variable $\mathbf{u}$ with mean $\mu_u$ and variance $\sigma_u^2$ is denoted by $\mathbf{u} \sim \mathcal{CN}\left(\mu_u,\sigma_u^2\right)$.

\section{System Model}\label{se2}
We consider a multi-band ISAC base station (BS) for weak-target detection, where the multi-band ISAC signal operates on $B$ non-continuous frequency bands based on carrier aggregation technology~\cite{wei2023carrier,liu2024carrier}, and the index set of multi-band is denoted by $\mathcal{B}=\left\{1,2,\cdots,B\right\}$.
In the detection period, let $N_\mathrm{c}^b$ subcarriers and $M_\mathrm{sym}^b$ OFDM symbols be utilized for each band $b\in\mathcal{B}$.
Accordingly, we denote the index sets for subcarriers and symbols as $\mathcal{N}_\mathrm{c}^b=\left\{1,2,\cdots,N_\mathrm{c}^b\right\}$ and $\mathcal{M}_\mathrm{sym}^b=\left\{1,2,\cdots,M_\mathrm{sym}^b\right\}$.
Future work will consider extending this framework to multi-target and MIMO scenarios.

\subsection{Multi-band ISAC Echo Signal Model}
For a target with range $R$ and radial velocity $v$,  the baseband echo signal received on subcarrier $n$ and symbol $m$ within band $b$ is given by 
\begin{equation}\label{eq1}
  \begin{aligned}
y_{b,n,m}=&\sqrt{P_\mathrm{t}^b}L_b\alpha_be^{-j2\pi n\Delta f_b \tau}e^{j2\pi mT_\mathrm{sym}^bf_b^\mathrm{d}}s_{b,n,m}\\ & + z_{b,n,m}, \quad b,n,m\in\left\{\mathcal{B},\mathcal{N}_\mathrm{c}^b, \mathcal{M}_\mathrm{sym}^b\right\}
  \end{aligned}  
\end{equation}
where $s_{b,n,m}$ denotes the known pilot symbol, with $\mathbb{E}\left[|s_{b,n,m}|^2\right]=1$ and $z_{b,n,m}\sim\mathcal{CN}\left(0,\sigma_{b,n,m}^2\right)$ is the additive white Gaussian noise;
$P_\mathrm{t}^b$ is the transmit power of band $b$ and $L_b=\sqrt{\frac{\lambda_b^2}{\left(4\pi\right)^3R^4}}$; 
$\lambda_b=\frac{c}{f_b}$ is the wavelength of band $b$, with $c$ and $f_b$ being the speed of light and the center frequency, respectively;
$\Delta f_b$ and $T_\mathrm{sym}^b$ represent the subcarrier spacing and symbol duration of band $b$, respectively;
$\tau=\frac{2R}{c}$ and $f_b^\mathrm{d}=\frac{2f_bv}{c}$ are the delay and Doppler shift of the target.

\subsection{Target Scattering Model}
To capture the wavelength dependent scattering of weak targets across wide span bands, we adopt a frequency selective Swerling II model, where the complex scattering coefficient $\alpha_b$ in \eqref{eq1} is modeled as $\alpha_b\sim\mathcal{CN}\left(0,\sigma_{\mathrm{rcs},b}^2\right), \forall b\in\mathcal{B}$~\cite{lv2025target}.
According to the Swerling II model~\cite{richards2005fundamentals}, the RCS is denoted as $\xi_b=|\alpha_b|^2$, which follows an exponential distribution with mean $\sigma_{\mathrm{rcs},b}^2=\mathbb{E}\left[\xi_b\right]$ and is jointly determined by the carrier frequency, aspect angle, and target geometry.

To characterize the frequency dependency, we consider a distributed target model with a projected length $d$.
According to \cite{Zhou2010Scattering}, the correlation coefficient $\rho_{i,j}$ between $\alpha_i$ and $\alpha_j$ is
\begin{equation}\label{eq2}
\begin{aligned}
   \rho_{i,j} = &J_0\left(2\pi\Delta f_{i,j}\cos\left(\theta_\mathrm{bis}\right)d/c\right)\\ & +J_2\left(2\pi\Delta f_{i,j}\cos\left(\theta_\mathrm{bis}\right)d/c\right), \quad i,j\in\mathcal{B}  
\end{aligned}
\end{equation}
where $J_0\left(\cdot\right)$ and $J_2\left(\cdot\right)$ are the zero-th and second-order Bessel functions of the first kind;
$\Delta f_{i,j} = |f_i-f_j|$ is the frequency difference;
$\theta_\mathrm{bis}=\frac{\theta_\mathrm{r}-\theta_\mathrm{t}}{2}$ is the bi-static angle, with $\theta_\mathrm{r}$ and $\theta_\mathrm{t}$ being the angles of arrival and departure, respectively.
Physically, the target's coherence bandwidth is defined as $B_\mathrm{c}\approx c/\left(2d\cos\theta_\mathrm{bis}\right)$, and we have the following reasonable assumptions.
\begin{itemize}
    \item \textbf{Intra-band correlation:} Within a single band, the frequency difference typically satisfies $\Delta f_{i,j}\ll B_\mathrm{c}$, leading to $2\pi\Delta f_{i,j}\cos\left(\theta_\mathrm{bis}\right)d/c\to 0$ and $\rho_{i,j\to 1}$.
    \item \textbf{Inter-band independence:} For non-contiguous bands in CA systems, the frequency gap is usually much larger than the coherence bandwidth (i.e., $\Delta f_{i,j}\gg B_\mathrm{c}$), which causes the correlation to vanish ($\rho_{i,j}\approx 0$)~\cite{Zhou2010Scattering}.
\end{itemize}
Consequently, the frequency bands are statistically independent and
\begin{equation}\label{eq3}
\mathbb{E}\left[\alpha_i\alpha_j^*\right] = \left\{\begin{matrix}
 \sigma_{\mathrm{rcs},i}^2 & i=j\\
 0 & i\neq j
\end{matrix}\right. 
\end{equation}

\section{GLRT-Based Performance
Characterization}\label{se3}
This section focuses on the detection performance derivation of multi-band ISAC systems and provides the motivation and theoretical basis for the optimization in Section \ref{se4}. 

\subsection{Joint Detection Statistic based on GLRT}\label{sec3-a}
When the target parameters ($\tau$, $f_b^\mathrm{d}$) are unknown, the global detector based on GLRT requires a grid search in a two-dimensional parameter space~\cite{Chowdhury2025ISAC}.
For candidate grid $(\tilde{\tau},\tilde{f}_b^\mathrm{d})$,
we perform coherent accumulation on a total of $N_\mathrm{c}^bM_\mathrm{sym}^b$ time-frequency resource elements within band $b$ to obtain the detection statistic $r_b$ for band $b$
\begin{equation}\label{eq4}
{\fontsize{9}{9}\begin{aligned}
   r_b & = \sum_m\sum_n y_{b,n,m}\left[s_{b,n,m}e^{-j2\pi n\Delta f_b \tilde{\tau}}e^{j2\pi mT_\mathrm{sym}^b\tilde{f}_b^\mathrm{d}}\right]^*
 \\ & = \underbrace{\sum_m\sum_n\sqrt{P_\mathrm{t}^b}L_b\alpha_be^{-j2\pi n\Delta f_b (\tau-\tilde{\tau})}e^{j2\pi mT_\mathrm{sym}^b(f_b^\mathrm{d}-\tilde{f}_b^\mathrm{d})}}_{\text{signal term } s_b} \\ & \quad + \underbrace{\sum_m\sum_n z_{b,n,m}\left[s_{b,n,m}e^{-j2\pi n\Delta f_b \tilde{\tau}}e^{j2\pi mT_\mathrm{sym}^b\tilde{f}_b^\mathrm{d}}\right]^*}_{\text{noise term } z_b}.
\end{aligned}}
\end{equation}
Therefore, the average signal-to-noise ratio (SNR) $\bar{\gamma}_b$ of $r_b$ can be derived as
\begin{equation}\label{eq5}
\bar{\gamma}_b = \frac{\mathbb{E}\left[|s_b|^2\right]}{\mathbb{E}\left[|z_b|^2\right]}= \frac{M_\mathrm{sym}^bN_\mathrm{c}^bL_b^2P_\mathrm{t}^b\sigma_{\mathrm{rcs},b}^2}{\sigma_{b,n,m}^2}.
\end{equation}

Since the frequency bands are statistically independent, the joint detection statistic based on GLRT is defined as~\cite{Chowdhury2025ISAC}
\begin{equation}\label{eq6}
\mathcal{L}=\underset{\tilde{\tau},\tilde{f}_b^\mathrm{d}}{\mathrm{max}} \frac{\prod_{b\in\mathcal{B}}p_{r_b}\left(x|\mathcal{H}_1\right)}{\prod_{b\in\mathcal{B}}p_{r_b}\left(x|\mathcal{H}_0\right)}
\mathop{\gtrless}_{\mathcal{H}_0}^{\mathcal{H}_1}\eta ,
\end{equation}
where $\eta$ denotes the threshold and $p(\cdot)$ is the probability density function (PDF).
Under both hypotheses, $r_b$ follows this distribution
\begin{equation}\label{eq7}
{\fontsize{8}{8}
\left\{\begin{matrix}
\mathcal{H}_0 : & r_b = z_b\sim\mathcal{CN}\left(0,M_\mathrm{sym}^bN_\mathrm{c}^b\sigma_{b,n,m}^2\right) \\
\mathcal{H}_1 : & r_b = s_b + z_b\sim\mathcal{CN}\left(0,M_\mathrm{sym}^bN_\mathrm{c}^b\sigma_{b,n,m}^2\left(\bar{\gamma_b}+1\right)\right)
\end{matrix}\right.}.
\end{equation}
We take the logarithm of both sides of \eqref{eq6} and obtain \eqref{eq8} shown in the top of this page.
\begin{figure*}
\begin{equation}\label{eq8}
{\fontsize{7}{7}
\begin{aligned}
  \ln{\mathcal{L}} & = \underset{\tilde{\tau},\tilde{f}_b^\mathrm{d}}{\mathrm{max}} \sum_{b\in\mathcal{B}}\left[\ln{p\left(r_b|\mathcal{H}_1\right)-\ln{p\left(r_b|\mathcal{H}_0\right)}}\right]\mathop{\gtrless}_{\mathcal{H}_0}^{\mathcal{H}_1}\ln{\eta}
  \\ &= \underset{\tilde{\tau},\tilde{f}_b^\mathrm{d}}{\mathrm{max}} \sum_{b\in\mathcal{B}}\left[\left(-\ln{\left(\pi M_\mathrm{sym}^bN_\mathrm{c}^b\sigma_{b,n,m}^2\left(\bar{\gamma}_b+1\right)\right)}-\frac{|r_b|^2}{M_\mathrm{sym}^bN_\mathrm{c}^b\sigma_{b,n,m}^2\left(\bar{\gamma}_b+1\right)}\right)-\left(-\ln{\left(\pi M_\mathrm{sym}^bN_\mathrm{c}^b\sigma_{b,n,m}^2\right)}-\frac{|r_b|^2}{M_\mathrm{sym}^bN_\mathrm{c}^b\sigma_{b,n,m}^2}\right)\right] \mathop{\gtrless}_{\mathcal{H}_0}^{\mathcal{H}_1}\ln{\eta}
  \\ &
  = \underset{\tilde{\tau},\tilde{f}_b^\mathrm{d}}{\mathrm{max}} \sum_{b\in\mathcal{B}}\left[\left(\frac{\bar{\gamma}_b}{\bar{\gamma}_b+1}\right)\frac{|r_b|^2}{M_\mathrm{sym}^bN_\mathrm{c}^b\sigma_{b,n,m}^2}+\ln\left(\frac{1}{\bar{\gamma}_b+1}\right)\right]\mathop{\gtrless}_{\mathcal{H}_0}^{\mathcal{H}_1}\ln{\eta}
  =\underset{\tilde{\tau},\tilde{f}_b^\mathrm{d}}{\mathrm{max}} \sum_{b\in\mathcal{B}}\left[\left(\frac{\bar{\gamma}_b}{\bar{\gamma}_b+1}\right)\frac{|r_b|^2}{M_\mathrm{sym}^bN_\mathrm{c}^b\sigma_{b,n,m}^2}\right]\mathop{\gtrless}_{\mathcal{H}_0}^{\mathcal{H}_1}\ln{\eta}+\sum_{b\in\mathcal{B}}\ln\left(\bar{\gamma}_b+1\right),
\end{aligned}}
\end{equation}
{\noindent} \rule[-10pt]{18cm}{0.1em}
\end{figure*}

Define the normalized energy statistic as $U_b = \frac{|r_b|^2}{M_\mathrm{sym}^bN_\mathrm{c}^b\sigma_{b,n,m}^2}$.
Eq.~\eqref{eq8} reveals that the GLRT-based detecter follows a linear weighted combination of $\{U_b\}_{b=1}^B$ with an optimal weight of $w_b^\mathrm{glrt}=\frac{\bar{\gamma}_b}{\bar{\gamma}_b+1}$.
In typical weak-target scenarios, $w_b^\mathrm{glrt}\approx\bar{\gamma}_b$~\cite{richards2005fundamentals}. Moreover, Appendix \ref{apA} proves that $\bar{\gamma}_b$ is the optimal weight under the deflection coefficient criterion. 
For criterion compatibility and low SNR consistency, we adopt $w_b=\bar{\gamma}_b$, simplifying \eqref{eq8} to
\begin{equation}\label{eq9}
   \Lambda= \underset{\tilde{\tau},\tilde{f}_b^\mathrm{d}}{\mathrm{max}} \sum_{b\in\mathcal{B}}w_bU_b\mathop{\gtrless}_{\mathcal{H}_0}^{\mathcal{H}_1}\mu
\end{equation}
where $\mu=\ln{\eta}+\sum\limits_{b\in\mathcal{B}}\ln\left(\bar{\gamma}_b+1\right)$ is the corrected threshold.

\subsection{Closed-form Detection Performance of Multi-band OFDM}
We define $X_b=w_bU_b=\bar{\gamma}_bU_b$ as a weighted detection component, and it is easy to see that $X_b\sim\mathrm{Exp}\left(\lambda_b\right)$.
Under both hypotheses, 
\begin{equation}\label{eq10}
 \mathcal{H}_0 : \lambda_{b|\mathcal{H}_0}=\frac{1}{\bar{\gamma}_b},\quad
 \mathcal{H}_1 :\lambda_{b|\mathcal{H}_1}=\frac{1}{\bar{\gamma}_b\left(\bar{\gamma}_b+1\right)}
\end{equation}
Observing \eqref{eq5}, we found that the differences in parameters across different non-continuous bands led to $\bar{\gamma}_i\neq \bar{\gamma}_j$ for $i\neq j$. 
This renders $\Lambda$ a sum of i.n.i.d. exponential random variables $X_b$. 
Therefore, we apply the characteristic function (CF) method~\cite{richards2005fundamentals}, and the derivations are as follows.

\subsubsection{False alarm probability}
Based on \eqref{eq10}, the CF of $\Lambda$ under the hypothesis $\mathcal{H}_0$ is 
\begin{equation}\label{eq11}
{\fontsize{9}{9}\begin{aligned}
   \varphi_0^\Lambda\left(t\right) &= \prod_{b=1}^B  \varphi_0^{X_b}\left(t\right)= \prod_{b=1}^B\int_{-\infty}^{\infty} e^{itx}p_{X_b}\left(x|\mathcal{H}_0\right)dx \\ & 
   = \prod_{b=1}^B\int_{0}^{\infty} \frac{1}{\bar{\gamma}_b}e^{-\left(\frac{1}{\bar{\gamma}_b}-it\right)x}dx =  \prod_{b=1}^B\frac{1}{1-\bar{\gamma}_bit},
\end{aligned}}
\end{equation}
which is a rational fraction of order $B$. 
We transform the product form into a summation form by partial fraction expansion $\varphi_0^\Lambda\left(t\right) = \sum_{b=1}^B\frac{A_b}{1-\bar{\gamma}_bit}$,
where 
{\fontsize{9}{9} \begin{equation}
 A_b = \lim_{it \to \frac{1}{\bar{\gamma}_b}}\left[\left(1-\bar{\gamma}_bit\right)\prod_{k=1}^B\frac{1}{1-\bar{\gamma}_kit}\right] = \prod_{k=1,k\neq b}\frac{\bar{\gamma}_b}{\bar{\gamma}_b-\bar{\gamma}_k}.
\end{equation}}

Then, the PDF of $\Lambda$ under $\mathcal{H}_0$ can be derived as
\begin{equation}\label{eq13}
{\fontsize{9}{9}\begin{aligned}
    p_{\Lambda}\left(x|\mathcal{H}_0\right) = \mathcal{F}^{-1}\left\{\sum\nolimits_{b=1}^B\frac{A_b}{1-\bar{\gamma}_b it}\right\} = \sum\nolimits_{b=1}^B\frac{A_b}{\bar{\gamma}_b}e^{-\frac{x}{\bar{\gamma}_b}},   
\end{aligned}}
\end{equation}
 and the closed-form expression of the probability of false alarm is 
{\fontsize{8}{8}\begin{equation}\label{eq14}
    P_\mathrm{FA}\left(\mu\right)= \int_{\mu}^{\infty} p_{\Lambda}\left(x|\mathcal{H}_0\right)dx=\sum_{b=1}^B\left[\prod_{k=1,k\neq b}^B\frac{\bar{\gamma}_b}{\bar{\gamma}_b-\bar{\gamma}_k}\right]e^{-\frac{\mu}{\bar{\gamma}_b}}.  
\end{equation}}

\subsubsection{Detection probability}
The derivation of the detection probability  $P_\mathrm{D}$ is similar to that of the false alarm probability, the only difference being
\begin{equation}\label{eq15}
   p_{X_b}\left(x|\mathcal{H}_1\right) = \frac{1}{\bar{\gamma}_b\left(\bar{\gamma}_b+1\right)}e^{-\frac{x}{\bar{\gamma}_b\left(\bar{\gamma}_b+1\right)}}.
\end{equation}

Therefore, the CF and PDF of $\Lambda$ under the hypothesis $\mathcal{H}_1$ are
\begin{equation}\label{eq16}
    \varphi_1^\Lambda\left(t\right) = \prod_{b=1}^B\frac{1}{1-\bar{\gamma}_b\left(\bar{\gamma}_b+1\right)it},
\end{equation}
and
\begin{equation}\label{eq17}
{\fontsize{7}{7}\begin{aligned}
&p_\Lambda\left(x|\mathcal{H}_1\right) =\sum_{b=1}^B\left[\prod_{k=1,k\neq b}^B\frac{\bar{\gamma}_b\left(\bar{\gamma}_b+1\right)}{\bar{\gamma}_b\left(\bar{\gamma}_b+1\right)-\bar{\gamma}_k\left(\bar{\gamma}_k+1\right)}\right]\frac{e^{-\frac{x}{\bar{\gamma}_b\left(\bar{\gamma}_b+1\right)}}}{\bar{\gamma}_b\left(\bar{\gamma}_b+1\right)}.  \end{aligned}}
\end{equation}
Based on \eqref{eq17}, the closed-form expression of the detection probability is 
{\fontsize{8}{8}\begin{equation}\label{eq18}
   P_\mathrm{D} = \sum_{b=1}^B\left[\prod_{k=1,k\neq b}^B\frac{\bar{\gamma}_b\left(\bar{\gamma}_b+1\right)}{\bar{\gamma}_b\left(\bar{\gamma}_b+1\right)-\bar{\gamma}_k\left(\bar{\gamma}_k+1\right)}\right]e^{-\frac{\mu}{\bar{\gamma}_b\left(\bar{\gamma}_b+1\right)}}. 
\end{equation}}
\textbf{Remark 1 (Single-band Baseline):}
\textit{To evaluate multi-band performance and robustness, a single-band signal with equivalent resources serves as a benchmark. Its closed-form is detailed in Appendix B and used in Section \ref{se5} to quantify multi-band advantages.}

\section{Joint Power and Time-Frequency Resource Allocation Scheme}\label{se4}
This section proposes a joint power and time-frequency resource allocation scheme to maximize the detection performance of the multi-band ISAC system. 
In practice, the proposed scheme is applicable to the following sensing framework: First, the system acquires noise variance through silence and obtains the prior average RCS of the weak target according to the target type. Based on this information, power and time-frequency resources are jointly optimized in the current sensing frame. 
Finally, the RCS and channel characteristics are updated based on the sensing results, and resources are optimized for the next sensing frame accordingly.

\subsection{Problem Formulation}\label{se4-A}
Building upon \eqref{eq18}, we formulate a joint optimization problem to maximize the detection probability $P_\mathrm{D}$.
Specifically, given the channel characteristics ($\left\{\sigma_{b,n,m}^2,\sigma_{\mathrm{rcs},b}^2\right\}_{b=1}^B$) and resource budgets ($P_\mathrm{total}=\sum\limits_{b=1}^BP_\mathrm{t}^b$, $S_\mathrm{total}=\sum\limits_{b=1}^BN_\mathrm{c}^bM_\mathrm{sym}^b$), the resource allocation problem is expressed as
\begin{subequations}
{\fontsize{9}{9} \begin{align}
\left(\text{P1}\right) & \quad \underset{\{P_b,S_b\}_{b=1}^B}{\max} 
\sum\limits_{b=1}^B \left[\begin{array}{l}
   \left(\prod\limits_{k=1,k\neq b}^B\frac{\varrho_b\left(\varrho_b+1\right)}{\varrho_b\left(\varrho_b+1\right)-\varrho_k\left(\varrho_k+1\right)}\right)\\ \times e^{-\frac{\mu}{\varrho_b\left(\varrho_b+1\right)}} \nonumber    
\end{array}\right] \\ &
\text{s.t.}\ \  \sum_{b=1}^BP_b\le P_\mathrm{total}, \vartheta_\mathrm{min}P_\mathrm{total}\le P_b\le \vartheta_\mathrm{max}P_\mathrm{total} \label{eq19a} \\ &
\quad \quad \sum_{b=1}^BS_b\le S_\mathrm{total},\ \vartheta_\mathrm{min}S_\mathrm{total}\le S_b \le \vartheta_\mathrm{max}S_\mathrm{total} \label{eq19b} \\ & \quad \quad S_b \in \mathbb{Z}^+, \forall b \label{eq19c},
\end{align} } 
\end{subequations}
where $\varrho_b=S_bP_b\Omega_b$ and $S_b = M_bN_b$.
Constraints \eqref{eq19a} and \eqref{eq19b} define the total budgets and per-band boundary limits for transmit power and time-frequency resources, respectively, where $\vartheta_\mathrm{min},\vartheta_\mathrm{max}\in\left(0,1\right]$ are scaling factors;
Constraint \eqref{eq19c} imposes positive integer constraints on $S_b$.
We do not consider the subcarrier and symbol subdivisions of each band for specific range or velocity limitations.
Since (P1) is NP-hard, we propose an ADMM-based algorithm to solve it.

\subsection{ADMM-based Algorithm}\label{sec4-B}
By introducing the auxiliary variable $E_b = S_b P_b$ and relaxing the integer constraint \eqref{eq19c} to $S_b\in\mathbb{R}_+$, (P1) becomes
\begin{subequations}
{\fontsize{9}{9} \begin{align}
\left(\text{P2}\right) & \underset{\{E_b,P_b,S_b\}_{b=1}^B}{\min} 
-\sum\limits_{b=1}^B \left[\begin{array}{l}
   \left(\prod\limits_{k=1,k\neq b}^B\frac{\varpi_b\left(\varpi_b+1\right)}{\varpi_b\left(\varpi_b+1\right)-\varpi_k\left(\varpi_k+1\right)}\right)\\ \times e^{-\frac{\mu}{\varpi_b\left(\varpi_b+1\right)}} \nonumber    
\end{array}\right] \\ &
\text{s.t.} \quad E_b-S_bP_b = 0,\forall b \label{eq20a} \\ &
\quad \quad \quad \eqref{eq19a}, \eqref{eq19b} \nonumber.
\end{align}  } 
\end{subequations}
Then, we derive an augmented Lagrangian from (P2)
\begin{equation}\label{eq21}
{\fontsize{9}{9}\begin{aligned}
L =& -F + \sum_{b=1}^B\lambda_b\left(E_b-S_bP_b\right) +\frac{\rho}{2} \sum_{b=1}^B\left(E_b-S_bP_b\right)^2,
\end{aligned}}
\end{equation}
where $\rho>0$ is the penalty parameter, $\left\{\lambda_b\right\}_{b=1}^B$ are the Lagrange multipliers, and 
\begin{equation}\label{eq22}
{\fontsize{7}{7}\begin{aligned}
   F = \sum\limits_{b=1}^B \left[
   \prod\limits_{k=1,k\neq b}^B\frac{\varpi_b\left(\varpi_b+1\right)}{\varpi_b\left(\varpi_b+1\right)-\varpi_k\left(\varpi_k+1\right)}\right] e^{-\frac{\mu}{\varpi_b\left(\varpi_b+1\right)}}.   
\end{aligned}}
\end{equation}
We divide \eqref{eq22} into three subproblems and adjust them by $\left\{\lambda_b\right\}_{b=1}^B$, where the steps of the $g$-th iteration are as follows.

\textit{\textbf{Step 1:}}(Update $\left\{E_b\right\}_{b=1}^B$ and fix $\left\{S_b^{(g)},P_b^{(g)},\lambda_b^{(g)}\right\}_{b=1}^B$) Given that the constraint \eqref{eq20a} is decoupled, the updates are parallelizable, with the update of $E_b$ is expressed as
\begin{equation}\label{eq23}
{\fontsize{8}{8}\begin{aligned}
 E_b^{(g+1)} = \underset{E_b}{\text{arg min}} \left[-F+\frac{\rho}{2}\sum_{b=1}^B\left(E_b-S_b^{(g)}P_b^{(g)}+\frac{\lambda_b^{(g)}}{\rho}\right)^2\right],   
\end{aligned}}  
\end{equation}
which is a non-convex problem and can be solved via the sequential quadratic programming (SQP) method~\cite{Boggs_Tolle_1995}.

\textit{\textbf{Step 2:}}(Update $\left\{S_b\right\}_{b=1}^B$ and fix $\left\{E_b^{(g+1)},P_b^{(g)},\lambda_b^{(g)}\right\}_{b=1}^B$)
The update of $\left\{S_b\right\}_{b=1}^B$ is expressed as
\begin{subequations}\label{eq24}
    \begin{align}
        \text{(P3)} & \underset{\left\{S_b\right\}_{b=1}^B}{\min} \sum_{b=1}^B\left(E_b^{(g+1)}-S_bP_b^{(g)}+\frac{\lambda_b^{(g)}}{\rho}\right)^2,\quad  
        \text{s.t.} \quad \eqref{eq19b} \nonumber
    \end{align}
\end{subequations}
which constitutes a linearly constrained QP solvable by the CVX tool~\cite{liu2025cooperative}. To recover integer feasibility, the obtained solution is rounded to the nearest positive integer.

\textit{\textbf{Step 3:}}(Update $\left\{P_b\right\}_{b=1}^B$, fix $\left\{E_b^{(g+1)},S_b^{(g+1)},\lambda_b^{(g)}\right\}_{b=1}^B$)
Similar to (P3), the update of $\left\{P_b\right\}_{b=1}^B$ is given by 
\begin{subequations}\label{eq25}
  {\fontsize{9}{9}  \begin{align}
        \text{(P4)} & \underset{\left\{P_b\right\}_{b=1}^B}{\min} \sum_{b=1}^B\left(E_b^{(g+1)}-S_b^{(g+1)}P_b+\frac{\lambda_b^{(g)}}{\rho}\right)^2,\quad  
        \text{s.t.} \quad \eqref{eq19a} \nonumber
    \end{align}}
\end{subequations}
which is solved via the CVX tool~\cite{liu2025cooperative}.

\textit{\textbf{Step 4:}}(Update $\left\{\lambda_b^{(g)}\right\}_{b=1}^B$)
The updated $\lambda_b$ is
\begin{equation}\label{eq26}
   \lambda_b^{(g+1)} = \lambda_b^{(g)} + \rho\left(E_b^{(g+1)}-S_b^{(g+1)}P_b^{(g+1)}\right). 
\end{equation}
The iterations from Step 1 to Step 4 continue until the following convergence criterion is met:
\begin{equation}\label{eq27}
 N_\mathrm{res} = \frac{\sqrt{\sum_{b=1}^B\left(E_b^{(g+1)}-S_b^{(g+1)}P_b^{(g+1)}\right)^2  }}{P_\mathrm{total}S_\mathrm{total}} \le \varepsilon_\mathrm{pri},
\end{equation}
where $N_\mathrm{res}$ and $\varepsilon_\mathrm{pri}$ denote the normalized residual and convergence tolerance, respectively. Upon convergence, the optimal parameters $\left\{P_b^*,S_b^*\right\}_{b=1}^B$ are obtained.

\section{Simulation Results}\label{se5}
This section validates our analytical derivations, highlights multi-band detection gains over a single-band baseline, and assesses the proposed joint resource allocation scheme. 
We consider four typical non-continuous bands at $\{2.6,3.5,26,28\}$ GHz~\cite{wei2023carrier,liu2024carrier}, and the system parameters are set as follows: 
The total numbers of subcarriers and symbols are 1024 and 64, respectively;
$P_\mathrm{total}=1024$ mW, $R=100$ m, and the average target RCS $\in[-20,0]$ dBsm (Swerling II)~\cite{richards2005fundamentals};
For ADMM optimization, we set $\rho=0.5$, $\varepsilon_\mathrm{pri}=0.001$, $\vartheta_\mathrm{min} =0.05$, and $\vartheta_\mathrm{max} =0.85$;
Performance is evaluated over 500 statistical experiments with $P_\mathrm{FA}\in\{10^{-12},10^{-8},10^{-4}\}$ on a device with an Intel i9-13900H CPU.

\subsection{Verification of Closed-form Derivations}\label{se5-A}
% The accuracy of the analytical derivation is validated by comparing the theoretical mean detection probability with that of the link-level, where the mean detection probability is defined as the average of the total statistical results. 
We validate the closed forms against link-level results using 5000 Monte Carlo trials per statistical experiment.

\begin{figure}
    \centering
    \includegraphics[width=0.30\textwidth]{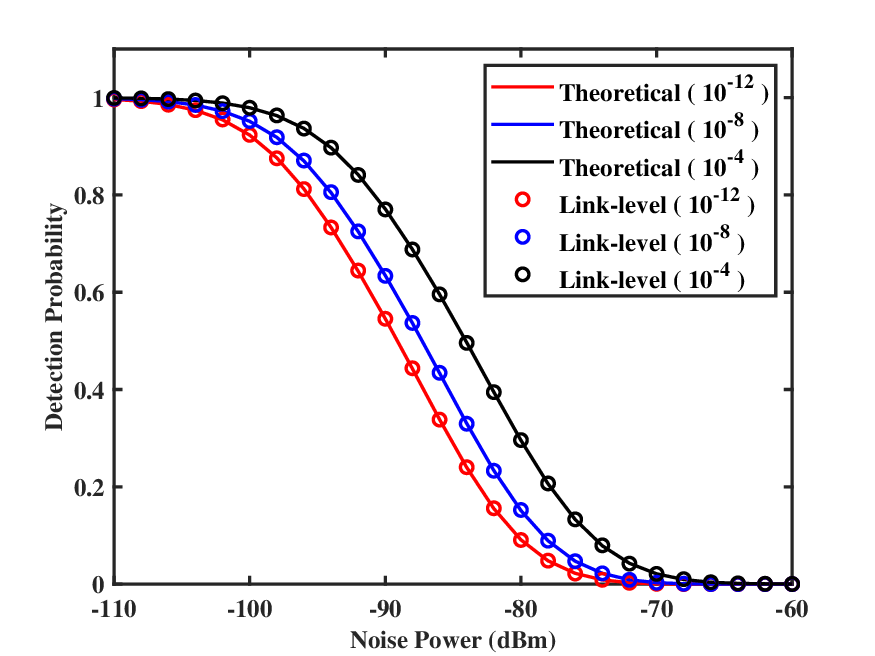}
    \caption{The mean detection probability of both the theoretical and link-level results}
    \label{fig1}
\end{figure}

As demonstrated in Fig.~\ref{fig1}, across various false alarm probabilities, the link-level results match the theoretical closed-form curves almost perfectly. 
Moreover, the detection probability decreases as the false alarm probability drops or the noise power increases, aligning with signal detection principles.
These observations verify our analytical derivations.

\subsection{Performance Gain of Multi-band}
In the simulation, the multi-band signal features a uniform allocation of time-frequency and power resources across all bands; meanwhile, the single-band baseline utilizes randomly selected frequencies to ensure a generalized benchmark.

\begin{figure}[!htbp]
    \centering
    \includegraphics[width=0.30\textwidth]{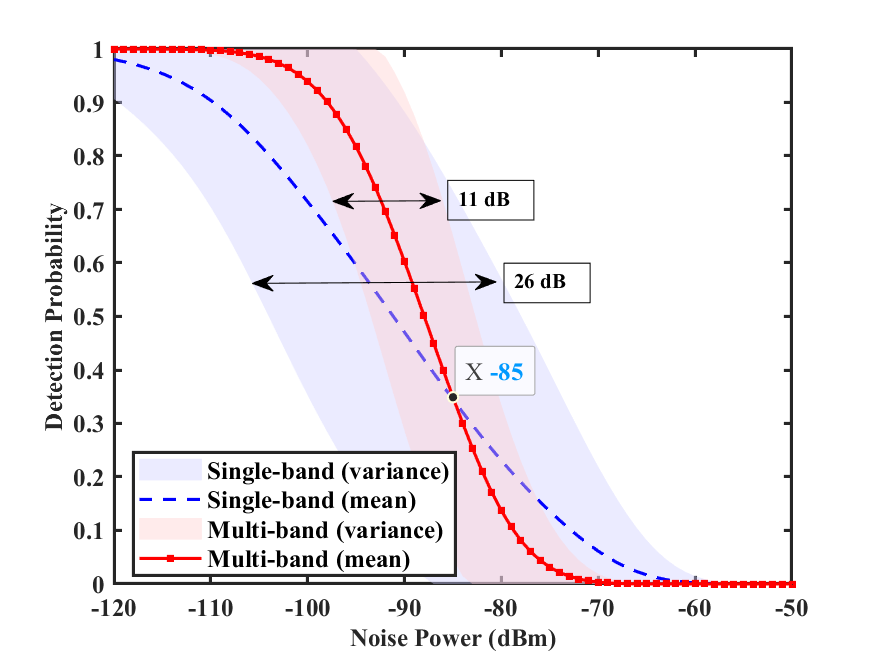}
    \caption{Multi-band signal \textit{v.s.} traditional single-band signal}
    \label{fig2}
\end{figure}

Fig.~\ref{fig2} illustrates the mean and variance of the detection probability at a false alarm probability of $10^{-8}$, where variance characterizes the robustness and mean characterizes the average performance. 
The multi-band signal outperforms its single-band counterpart when the noise power is below -83 dBm. 
More importantly, the variance results indicate that the performance fluctuation of the multi-band signal is approximately 10 dB, which is significantly narrower than the 30 dB variance observed in the single-band case. 
This confirms that joint processing across non-contiguous bands provides frequency diversity to mitigate fading-induced degradation, offering superior stability.

\subsection{ADMM-based Joint Optimization Method}
\begin{figure}[!htbp]
    \centering  \includegraphics[width=0.30\textwidth]{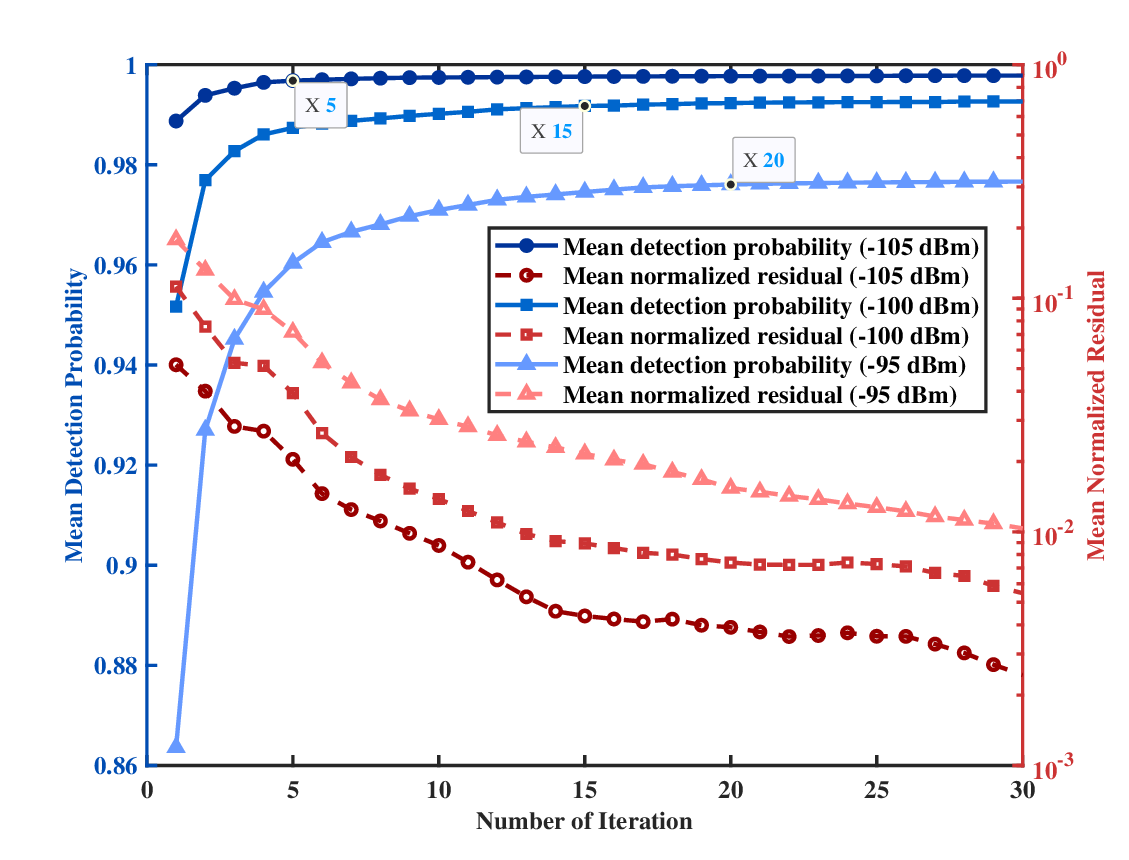}
    \caption{The mean detection probability and the mean normalized residual with noise power of $\{-105,-100,-95\}$ dBm}
    \label{fig3}
\end{figure}

To evaluate the convergence of the proposed algorithm, we analyze the mean detection probability and the mean normalized residual over iterations under varying noise power levels. 
As illustrated in Fig.~\ref{fig3}, the mean detection probability increases and eventually plateaus within a limited number of iterations, while the mean normalized residual steadily declines. 
Higher noise power requires more iterations to reach a steady state, providing practical guidance for configuring the iteration limit of the algorithm. 
These results demonstrate the stability and efficiency of the proposed ADMM-based algorithm.

\begin{figure}[!htbp]
    \centering  \includegraphics[width=0.30\textwidth]{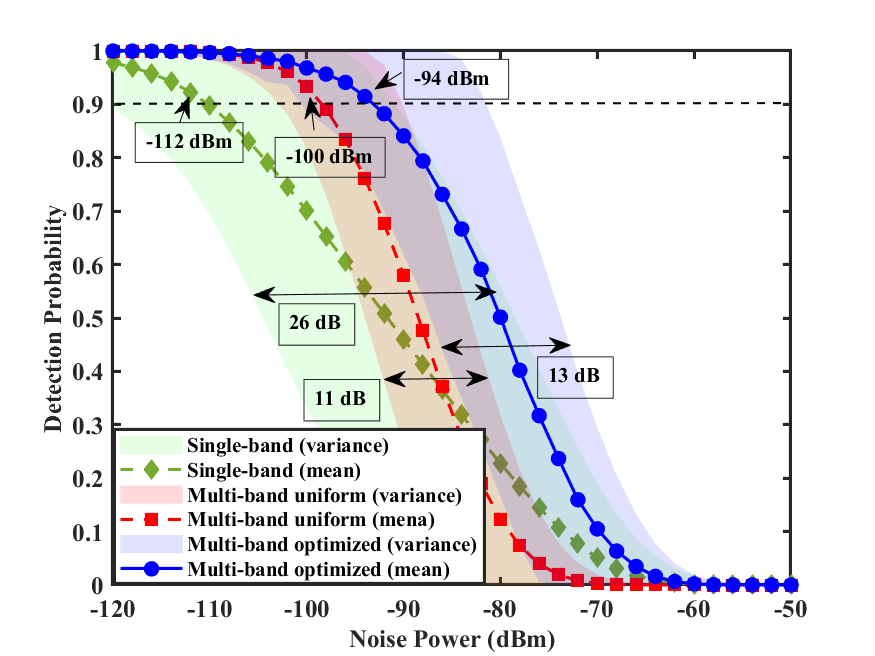}
    \caption{Optimized multi-band signal \textit{v.s.} uniform multi-band signal \textit{v.s.} single-band signal}
    \label{fig4}
\end{figure}

Then, the detection performance of the optimized multi-band signal is compared with the uniform multi-band signal and the single-band baseline. 
Fig.~\ref{fig4} illustrates the following phenomena.
\begin{itemize}
    \item The optimized multi-band signal outperforms the single-band baseline in both mean performance and upper limits because joint optimization offers more degrees of freedom to exploit the nonlinear characteristics of the detection probability.
    \item At a 90\% detection probability, the optimized multi-band signal yields 6 dB and 18 dB detection gains over uniform multi-band and single-band signals respectively, owing to the multi dimensional optimization of power and time-frequency resources that fully exploits heterogeneous band characteristics.
    \item The variances for the three signals are 13 dB, 11 dB, and 26 dB. Although the optimized signal shows a slightly higher variance than the uniform signal due to the dynamic reconfiguration of resources, its robustness remains significantly superior to the single-band baseline.
\end{itemize}
In summary, joint optimization enhances detection gain and robustness in multi-band ISAC systems.


\section{Conclusion}\label{se6}
Multi-band ISAC has gained significant interest owing to its capability to provide frequency diversity for robust detection. 
In this paper, closed-form expressions for detection and false alarm probabilities were derived based on GLRT. 
To fully exploit the multi-band potential, a joint power and time-frequency resource allocation scheme was developed using a proposed ADMM-based algorithm. Simulations validated the theoretical derivations and demonstrated the substantial detection gains and superior robustness of multi-band signals compared to single-band counterparts.
% This paper provides a reference for MPCs-aided MIMO-OFDM ISAC research.

\begin{appendices}
\section{Proof of the Optimal Weights}\label{apA}
Under the deflection coefficient criterion~\cite{Picinbono1995Deflection}, the deflection coefficient of the joint detection statistic is defined as
\begin{equation}\label{ape1}
    D(\Lambda) = \left( \mathbb{E}[\Lambda|\mathcal{H}_1] - \mathbb{E}[\Lambda|\mathcal{H}_0] \right)^2/\text{Var}[\Lambda|\mathcal{H}_0].  
\end{equation}
Based on the distributions in \eqref{eq7}, the moments of $U_b$ are
\begin{equation}\label{ape2}
   \mathbb{E}[U_b|\mathcal{H}_0] = \text{Var}[U_b|\mathcal{H}_0] = 1, \quad \mathbb{E}[U_b|\mathcal{H}_1] = \bar{\gamma}_b + 1 .
\end{equation}
Substituting \eqref{ape2} into \eqref{ape1} yields
\begin{equation}\label{ape3}
   D(\Lambda) = \left( \sum\limits_{b\in \mathcal{B}} w_b \left(\bar{\gamma}_b+1\right)-\sum\limits_{b\in \mathcal{B}} w_b \right)^2/\sum\limits_{b\in \mathcal{B}} \left(w_b\right)^2. 
\end{equation}
By the Cauchy-Schwarz inequality, $\left(\sum\limits_{b\in \mathcal{B}}w_b\bar{\gamma}_b\right)^2\le \sum\limits_{b\in \mathcal{B}}\left(w_b\right)^2\sum\limits_{b\in \mathcal{B}}\left(\bar{\gamma}_b\right)^2$, with equality holding if and only if $w_b=k\bar{\gamma}_b$ for $k>0$. Since the detection performance is invariant to the constant scaling of $\Lambda$, the optimal weight is $w_b=\bar{\gamma}_b$.

\section{Detection Performance for Single-band OFDM}
We assume a single-band signal with total time-frequency $\sum_{b=1}^BN_\mathrm{c}^bM_\mathrm{sym}^b$ and transmit power $\sum_{b=1}^BP_\mathrm{t}^b$ resources concentrated in band $s\in\mathcal{B}$. 
The distributions of the joint detection statistic under the two hypotheses are then
\begin{equation}\label{ape4}
{\fontsize{8}{8}\left\{\begin{matrix}
 \mathcal{H}_0 :& r_\mathrm{total}\sim\mathcal{CN}\left(0,\sum\limits_{b=1}^BN_\mathrm{c}^bM_\mathrm{sym}^b\sigma_{s,n,m}^2\right)\\
 \mathcal{H}_1: &r_\mathrm{total}\sim\mathcal{CN}\left(0,\sum\limits_{b=1}^BN_\mathrm{c}^bM_\mathrm{sym}^b\sigma_{s,n,m}^2\left(\bar{\gamma}_\mathrm{total}+1\right)\right)
\end{matrix}\right..} 
\end{equation}
where 
\begin{equation}\label{ape5}
    \bar{\gamma}_\mathrm{total} = \left( \sum\limits_{b=1}^BN_\mathrm{c}^bM_\mathrm{sym}^b\sum\limits_{b=1}^BP_\mathrm{t}^bL_s^2\sigma_{\mathrm{rcs},s}^2\right)/\sigma_{s,n,m}^2.
\end{equation}
Therefore, the joint detection statistic of single-band signal based on GLRT is defined as
\begin{equation}\label{ape6}
    \mathcal{L}_s = \underset{\tilde{\tau},\tilde{f}_s^\mathrm{d}}{\max}\frac{p_{r_\mathrm{total}}\left(x|\mathcal{H}_1\right)}{p_{r_\mathrm{total}}\left(x|\mathcal{H}_0\right)}\mathop{\gtrless}_{\mathcal{H}_0}^{\mathcal{H}_1}\mu
\end{equation}

Define a normalized energy statistic of the single-band signal as $U_s = \frac{|r_\mathrm{total}|^2}{\sum\limits_{b=1}^BN_\mathrm{c}^bM_\mathrm{sym}^b\sigma_{s,n,m}^2}$, and 
\begin{equation}\label{ape7}
\mathcal{H}_0:  U_s\sim\mathrm{Exp}\left(1\right),\quad
  \mathcal{H}_1: U_s \sim\mathrm{Exp}\left(1+\bar{\gamma}_\mathrm{total}\right)
\end{equation}
Based on \eqref{ape7}, the detection ($P_{\mathrm{D},s}$) and false alarm ($P_{\mathrm{FA},s}$) probabilities of the single-band signal are
\begin{equation}\label{ape8}
   P_{\mathrm{FA},s} = \int_{\mu}^\infty p_{U_s}\left(x|\mathcal{H}_0\right) dx = e^{-\mu},
\end{equation}
and
\begin{equation}\label{ape9}
   P_{\mathrm{D},s} = \int_\mu^\infty p_{U_s}\left(x|\mathcal{H}_1\right)dx = e^{\left(\frac{-\mu}{\bar{\gamma}_\mathrm{total} +1}\right)}.
\end{equation}

\end{appendices}

\section*{Acknowledgment}
This work was supported in part by the Fundamental Research Funds for the Central Universities under Grant 2023RC18, in part by the National Key Research and Development Program of China under Grant 2020YFA0711302, in part by the Fundamental Research Funds for the Central Universities under Grant 2024ZCJH01, and in part by the National Natural Science Foundation of China (NSFC) under Grant 62271081, and U21B2014. 

% reference
\bibliographystyle{IEEEtran}
\bibliography{reference}

\end{document}